\documentclass[iop,apjl]{emulateapj}
\usepackage{CJKutf8}

\newcommand{\simless}{\mathbin{\lower 3pt\hbox
      {$\rlap{\raise 5pt\hbox{$\char'074$}}\mathchar"7218$}}} 
\newcommand{\simgreat}{\mathbin{\lower 3pt\hbox
     {$\rlap{\raise 5pt\hbox{$\char'076$}}\mathchar"7218$}}} 

\slugcomment{\today}

\shorttitle{ALMA and ngVLA observations of disks}
\shortauthors{Ricci et al.}

\begin{document}
\begin{CJK*}{UTF8}{gbsn}


\title{Investigating the early evolution of planetary systems with ALMA and the Next Generation Very Large Array}


\author{Luca Ricci}

\affil{Department of Physics and Astronomy, Rice University, 6100 Main Street, 77005 Houston, TX, USA; Department of Physics and Astronomy, California State University Northridge, 18111 Nordhoff Street, Northridge CA 91130, USA; Jet Propulsion Laboratory, California Institute of Technology, 4800 Oak Grove Drive, Pasadena, CA, 91109, USA; luca.ricci@rice.edu } 

\and

\author{Shang-Fei Liu (刘尚飞), Andrea Isella}

\affil{Department of Physics and Astronomy, Rice University, 6100 Main Street, 77005 Houston, TX, USA; shangfei.liu@gmail.com; isella@rice.edu } 

\and

\author{Hui Li (李晖)}
\affil{Theoretical Division, Los Alamos National Laboratory, 87545 Los Alamos, NM, USA; hli@lanl.gov}


\begin{abstract}

We investigate the potential of the Atacama Large Millimeter/submillimeter Array (ALMA) and the Next Generation Very Large Array (ngVLA) to observe substructures in nearby young disks which are due to the gravitational interaction between disk material and planets close to the central star. We simulate the gas and dust dynamics in the disk using the LA-COMPASS hydrodynamical code. We generate synthetic images for the dust continuum emission at sub-millimeter to centimeter wavelengths and simulate ALMA and ngVLA observations. We explore the parameter space of some of the main disk and planet properties that would produce substructures that can be visible with ALMA and the ngVLA. We find that ngVLA observations with an angular resolution of 5 milliarcsec at 3 mm can reveal and characterize gaps and azimuthal asymmetries in disks hosting planets with masses down to $\approx~5 M_{\oplus}$ $\approx 1 - 5$ au from a Solar-like star in the closest star forming regions, whereas ALMA can detect gaps down to planetary masses of $\approx 20~M_{\oplus}$ at 5 au. Gaps opened by super-Earth planets with masses $\approx 5 - 10 M_{\oplus}$ are detectable by the ngVLA in the case of disks with low viscosity ($\alpha \sim 10^{-5}$) and low pressure scale height ($h \approx 0.025$ au at 5 au). The ngVLA can measure the proper motion of azimuthal asymmetric structures associated with the disk-planet interaction, as well as possible circumplanetary disks on timescales as short as one to a few weeks for planets at $1 - 5$ au from the star.

\end{abstract}



\section{Introduction}
\label{sec:intro}

Young pre-Main Sequence (PMS) stars and brown dwarfs are surrounded by disks which are considered to be the cradles of planets \citep[see][for a recent review]{Andrews:2015}. 
The annular gaps discovered by ALMA in the HL Tau, TW Hya, HD 163296, and HD 169142 protoplanetary disks \citep[][]{Alma:2015,Andrews:2016,Isella:2016,Fedele:2017} are reminiscent of the interaction between newborn planets and the circumstellar material (Bryden et al. 1999). The comparison with theoretical models indicates that these structures might indeed result from the gravitational interaction between the circumstellar disk and giant planets with masses similar to those of Saturn or Jupiter orbiting the star at tens of au \citep[e.g.,][]{Dipierro:2015,Jin:2016,Isella:2016,Liu:2017}. If confirmed, this would imply that giant planets form at much larger orbital radii than postulated by current planet formation models.

The same observations also revealed that the mm-wave dust continuum emission arising within $\sim 20-30$ au from the star in relatively massive young disks is optically thick. The large optical depth prevents us from measuring the dust density and, therefore, image planet-driven density perturbations. In practice, ALMA observations cannot probe planets orbiting in these innermost regions, unless they are massive enough to carve very deep, optically thin, gaps in the dust distribution. ALMA observations might therefore miss the bulk population of forming planets, which likely have masses well below the mass of Jupiter and orbit at less than 10 au from the host star~\citep[see, e.g.,][]{Burke:2015}.

A natural solution to this problem consists in imaging protoplanetary disks at frequencies between $\approx 10 - 100$ GHz, where the dust continuum emission from the innermost disk regions is optically thin, but still bright enough to be detected. These wavelengths are covered by the Karl G. Jansky Very Large Array (VLA), which, however, lacks the angular resolution and sensitivity to efficiently search for signatures of planets in the innermost and densest disk regions.

Since the past few years, the astronomy community has initiated discussion for a future large area mm/radio array optimized for imaging of thermal emission to milli-arcsecond (mas) scales that will open new discovery space for studies of protoplanetary disks, among other subjects\footnote{For more information, see \texttt{https://science.nrao.edu/futures/ngvla}.}. Although the final design of the array has yet to be established, this Next Generation Very Large Array (ngVLA) is currently envisioned to observe at frequencies $\sim 1.2 - 116$ GHz, and to include $\sim$ 10$\times$ the collecting area of the VLA \& ALMA, $\sim$ 10$\times$ longer baselines (300 km) that would yield mas-resolution at the highest frequencies. 

The vastly larger collecting area, resolving power, and image quality of the ngVLA will transform the study of planet formation. A preliminary investigation shows that the ngVLA will be capable of detecting perturbations from Saturn mass planets orbiting the star at 5 au~\citep[][]{Isella:2015}. However, the scope of this initial study was limited to one particular case. 
For example, \citet{Isella:2015} did not explore the range of planet parameters (masses and orbital radii) that can be uniquely probed by the ngVLA, nor tried to optimize the array configuration and collecting area to maximize the scientific output of such research. Radio observations with the ngVLA can also detect emission from circumplanetary disks, the sites of satellite formation \citep{Zhu:2017}.


In this work, we expand the initial analysis discussed in \citet{Isella:2015} to investigate the impact of the unique mapping capabilities of the ngVLA on the study of planet formation in nearby star forming regions. We test the visibility of disk substructures due to the gravitational interaction with planets with masses down to the super-Earth regime. For this study, we vary some of the key properties of the disk and planets to explore the range of parameters that can be probed by the ngVLA \footnote{All the hydrodynamic models presented in this work can be downloaded from \texttt{https://github.com/shangfei/ngVLA}.}. In doing this, we also produce synthetic images with ALMA for all our models. This allows us to quantify the sensitivity of ALMA to disk substructures generated by the interaction with a planet at orbital separations $\simless~10$ au from the star, as well as to compare it with the sensitivity of the ngVLA.  

Section~\ref{sec:method} describes the method adopted for our investigation, such as the hydrodynamical code used to generate synthetic images for the dust continuum emission at multiple wavelengths, and the method used to simulate ngVLA and ALMA observations of our models. Section~\ref{sec:results} outlines the main results of our work. Discussion and conclusions are described in Sections~\ref{sec:discussion} and \ref{sec:conclusions}, respectively.

\section{Method}
\label{sec:method}

In this Section we describe the methodology adopted to generate the synthetic images of disks at sub-mm and radio wavelengths and to simulate ALMA and ngVLA observations at high angular resolution. 

\subsection{Hydrodynamic simulations}
\label{sec:hydro}

We used the LA-COMPASS (Los Alamos COMPutional Astrophysics Simulation Suite) bi-fluid hydrodynamic code \citep{Li:2005,Li:2009} to simulate the dynamics of gas and dust in circumstellar disks with one or more planets. This code accounts for the mutual aerodynamic interaction between gas and dust grains with different sizes, and is therefore capable of deriving predictions for the continuum emission from dust grains at multiple wavelengths. To treat the dust dynamics, we adopted the one-fluid algorithm described in \citet{Laibe:2014}. Simulations are in the 2D $\{r, \phi\}$ plane.

For the initial surface density of gas, we adopted the following radial profile:

\begin{equation}\label{eq:surface_density}
\Sigma_{\rm{gas}} (r , t = 0) = \Sigma_c \left( \frac{r}{r_c} \right)^{-\gamma} \exp{\left[-\left(\frac{r}{r_c}\right)^{2-\gamma}\right]},
\end{equation}
which is the self-similar solution for the evolution of viscous accretion disks, in the case of viscosity $\nu(r) \propto r^{\gamma}$ \citep{Lynden-Bell:1974}. 
In all the simulations, we adopted values of $\gamma = 0.8$ and $r_c = 17$ au. The value of the normalization constant $\Sigma_c$ was set by fixing the initial total gas mass in the disk, as described below. As we are interested in investigating structures in the disk generated by the gravitational interaction with one or more planets at orbital radii $< 10$ au, we truncated our disk models at an outer radius of 60 au. 

For the initial surface density of dust, we assumed:

\begin{equation}
\Sigma_{\rm{dust}} (r , t = 0) = 0.01 \times \Sigma_{\rm{gas}} (r , t = 0),
\end{equation}
where 0.01 is the dust-to-gas mass ratio in the Interstellar Medium.
Since the aerodynamic coupling with gas depends on the size of solids, we then distributed this total dust surface density in 11 bins in solid size, ranging between 0.007 mm and 1 cm. In each bin, the mass of dust particles was assigned by assuming a power-law index of $q = 3.5$ for the grain size distribution $n(a) \propto a^{-q}$ \citep[e.g.,][]{Ricci:2010}. With this grain size distribution, smaller grains carry most of the opacity at optical and infrared wavelengths, but they have negligible mass relative to largest grains. For the volume density of each solid, we assumed a value of $\rho_{\rm{solid}} = $ 1.2 g/cm$^{3}$, consistent with the dust model described in the next Section.

The mass of the star is $M_\star = 1~M_{\odot}$.
For our reference model, the pressure scale height in the disk is given by $h/r = 0.05 \times (r / 5\rm{au})^{0.25}$. In Section~\ref{sec:low_h} we discuss the effects of varying the pressure scale height in the disk, by running simulations with scale heights that are \textit{lower} by a factor of 2 than in our reference model.

Throughout the disk the local sound speed $c_s = c_s (r)$ is derived by imposing vertical hydrostatic equilibrium, i.e. $c_s = v_K \times h/r$, with $v_K$ being the local Keplerian rotation velocity.
The temperature derived from this sound speed well approximates  the midplane temperature computed using the Monte Carlo radiative transfer code RADMC-3D \citep{Dullemond:2012} for a disk around a PMS star with a mass of 1 $M_{\odot}$, a luminosity of 1 $L_{\odot}$, and an effective temperature of 5800 K, and under the assumption of hydrostatic equilibrium between gas pressure and stellar gravity.  

For the gas viscosity, we adopt the $\alpha$ prescription of \citet{Shakura:1973}. As described in Section~\ref{sec:results}, gas viscosity has the effect of washing out azimuthally asymmetric structures in the gas distribution, as well as of closing the gap otherwise opened by a planet and of diffusing dust particles into the gap. Hence, simulations with different values of $\alpha$ can generate disks with different morphologies. For this reason, in this work we consider two values of the $\alpha$ parameter: $10^{-3}$ (Sect.~\ref{sec:high_visc}) and $10^{-5}$ (Sect.~\ref{sec:low_visc}), which are consistent with recent observational constraints from the imaging of molecular lines with ALMA~\citep{Flaherty:2017}. 

In order to test the observability of disk substructures due to the interaction with one or more embedded planets, we considered different values for the planetary masses and orbital radii. The planetary masses are 300~$M_{\oplus}$ (similar to Jupiter), 100~$M_{\oplus}$ (similar to Saturn), 50~$M_{\oplus}$, 30~$M_{\oplus}$, 17~$M_{\oplus}$ (similar to Neptune), 10~$M_{\oplus}$, 5~$M_{\oplus}$, 2~$M_{\oplus}$ and 1~$M_{\oplus}$. The planets orbital radii are 1, 2.5 and 5 au. Planetary masses and orbital radii are kept fixed during each simulation. 

The size of the grid cells in our simulations was determined from a trade-off between having enough numerical resolution to well resolve the local pressure scale height and a reasonable usage of CPU time. We used 1536 grid cells in the radial direction and 1024 in the azimuthal direction. In order to minimize possible artificial boundary effects, we adopted an inner radius of $0.1 \times r_p$, where $r_p$ is the planet orbital radius.   

All the simulations were run for 1500 planet orbital periods. All the images shown in this paper refer to the configuration derived at 1500 orbits after the beginning of the simulation, corresponding to 1500 and 16700 years at the orbital radii of 1 and 5 au, respectively. 

\begin{table*}[htb!]
\centering
\begin{tabular}{ccccccc}
\hline
\hline
Planet Mass [$M_{\oplus}$] & \multicolumn{5}{c}{Planet orbital radius [au]} \vspace{1mm}\\
     &  \multicolumn{4}{c}{$h/r = 0.05 \times (r/5\rm{au})^{0.25}$} & $h/r = 0.025 \times (r/5\rm{au})^{0.25}$ & $h/r = 0.0125 \times (r/5\rm{au})^{0.25}$ \vspace{1mm}\\
     &  \multicolumn{2}{c}{$M_{\rm{disk}} = 8\times10^{-3}~M_{\odot}$} &  \multicolumn{2}{c}{$M_{\rm{disk}} = 8\times10^{-2}~M_{\odot}$} & $M_{\rm{disk}} = 4\times10^{-2}~M_{\odot}$ & $M_{\rm{disk}} = 4\times10^{-2}~M_{\odot}$ \vspace{1mm}\\
     & $\alpha=10^{-5}$~~ & $\alpha=10^{-3}$~~  & $\alpha=10^{-5}$~~  & $\alpha=10^{-3}$ & $\alpha=10^{-5}$ & $\alpha=10^{-5}$ \\

\hline
300 ($\approx$ Jupiter)      &  1, 5    &  1, 5   & 1, 5    & 1, 5  &    ...  &    ...     \\
100 ($\approx$ Saturn)     &  1, 5    &  1, 5   & 1    &    ...    &    ...  &    ...   \\
50        &    ...  &   5  & ...    &     ...    &    ... &    ...   \\
30        &     ... & 2.5, 5    & ...     &     ...   &    ...  &    ...   \\
17  ($\approx$ Neptune)      &   5   &  5   &  ...   &    ...    &    ... &    ...    \\
10        &   5   &   ...  &  ...   &   ...    &    5   &    ...   \\
5          &   2.5, 5   &   ...  &  ...   &   ...    & 5 &    ...   \\
2          &   ...   &   ...  &  ...   &   ...    & 1, 5 &    ...   \\
1          &   ...   &   ...  &  ...   &   ...   & ... & 1, 5    \\

\hline
\end{tabular}
\caption{Table showing the values of the disk and planet model parameters investigated by this study}
  \label{tab:myfirsttable}
\end{table*}

\subsection{Simulations of the ALMA and ngVLA observations}
\label{sec:hydro}

The outputs of the LA-COMPASS simulations described above were used to generate synthetic images for the continuum emission at different wavelengths observable with ALMA and the ngVLA. 
In particular, the density and temperature structures computed by our LA-COMPASS simulations were used as inputs for the ray-tracing algorithm in RADMC-3D. In particular, for the reference model described above the radial profile of the dust temperature is: $T_{d}(r) = 37$ K $\times (r/\rm{10au})^{-0.5}$.

At any given wavelength, the surface brightness map accounts for the emission from dust with different sizes.
For the grain size-dependent dust absorption coefficients we used the model adopted in \citet{Ricci:2010}. This consists in a simplified version of the \citet{Pollack:1994} models for porous composite spherical grains made of astronomical silicates, carbonaceous materials and water ice. The optical constants of the single constituents are combined together using the Bruggeman effective medium theory \citep{Bruggeman:1935}.  

For each simulation, we calculated synthetic images at wavelengths of 0.35, 0.87, 1.3, 3.0 mm, and 1 cm. The former three wavelengths were used for the ALMA observations, while the latter two for the ngVLA\footnote{Note that also ALMA can observe at $\lambda = 3.0$ mm, but the ALMA sensitivity and angular resolution at this wavelength are significantly worse than for the ngVLA, and for this reason we did not compute ALMA observations at that wavelength.}. In order to obtain a fair comparison of the imaging capabilities, we adjusted the declination of the disks to be $-24^{\circ}$ for ALMA, and $+24^{\circ}$ for the ngVLA. These correspond to the declinations of the Ophiuchus and Taurus nearby star forming regions, respectively.

We used the \texttt{SIMOBSERVE} task within the Common Astronomy Software Application (CASA) package to convert the synthetic images into visibility datasets in the Fourier domain. 
For the array configuration of the ALMA observations we used the file \texttt{alma.out28.cfg} available in the CASA package, which contains the longest 16 km baselines\footnote{In the case of the ALMA simulations at 0.35 mm, a significant fraction of the extended emission was filtered out by this extended array configuration. In that case, we also included observations with a more compact array configuration (file \texttt{alma.out10.cfg}) to recover the entire emission from the model.}. For the ngVLA simulations we considered a fiducial configuration made of 300 antennas, with 30\% of the antennas within a 1 km radius and maximum baselines of 300 km \citep{Carilli:2015,Carilli:2016}.

The \textit{(u,v) points sampled by the} simulated observations correspond to a 8-hour synthesis centered on transit. In the ngVLA case, we did not use \texttt{SIMOBSERVE} to set the noise. Following the method outlined by \citet{Carilli:2016} to add thermal noise in ngVLA datasets, we used the \texttt{SIMNOISE} task to assign a noise level per visibility. This was set to give a rms level of 0.5 and 0.15 $\mu$Jy/beam on the final continuum maps at 3 mm and 1 cm, respectively, for Briggs weighting with robust parameter R$=$-1.  
Adopting the ngVLA nominal parameters listed in Table 1 of \citet{Carilli:2015}, these rms values correspond to on-time integrations of about 20 hours.  

Imaging was performed using the \texttt{CLEAN} task in CASA. Unless stated otherwise, the ALMA images were computed with a Briggs weighting with robust parameter R$=$-2 (uniform weighting), while for the ngVLA we chose R$=$-1. These values generally optimize the visibility on the final maps of the structures predicted by the models.   
We also employed a multiscale clean approach to better recover both compact emission at high brightness, and large, diffuse structures in the models.

\section{Results}
\label{sec:results}

In this Section we present the results of the LA-COMPASS simulations for disks with embedded planets and the predictions for interferometric observations with ALMA and the ngVLA. 

\subsection{Models with $\alpha = 10^{-3}$}
\label{sec:high_visc}

\begin{figure*}
\centering
\includegraphics[scale=1.0]{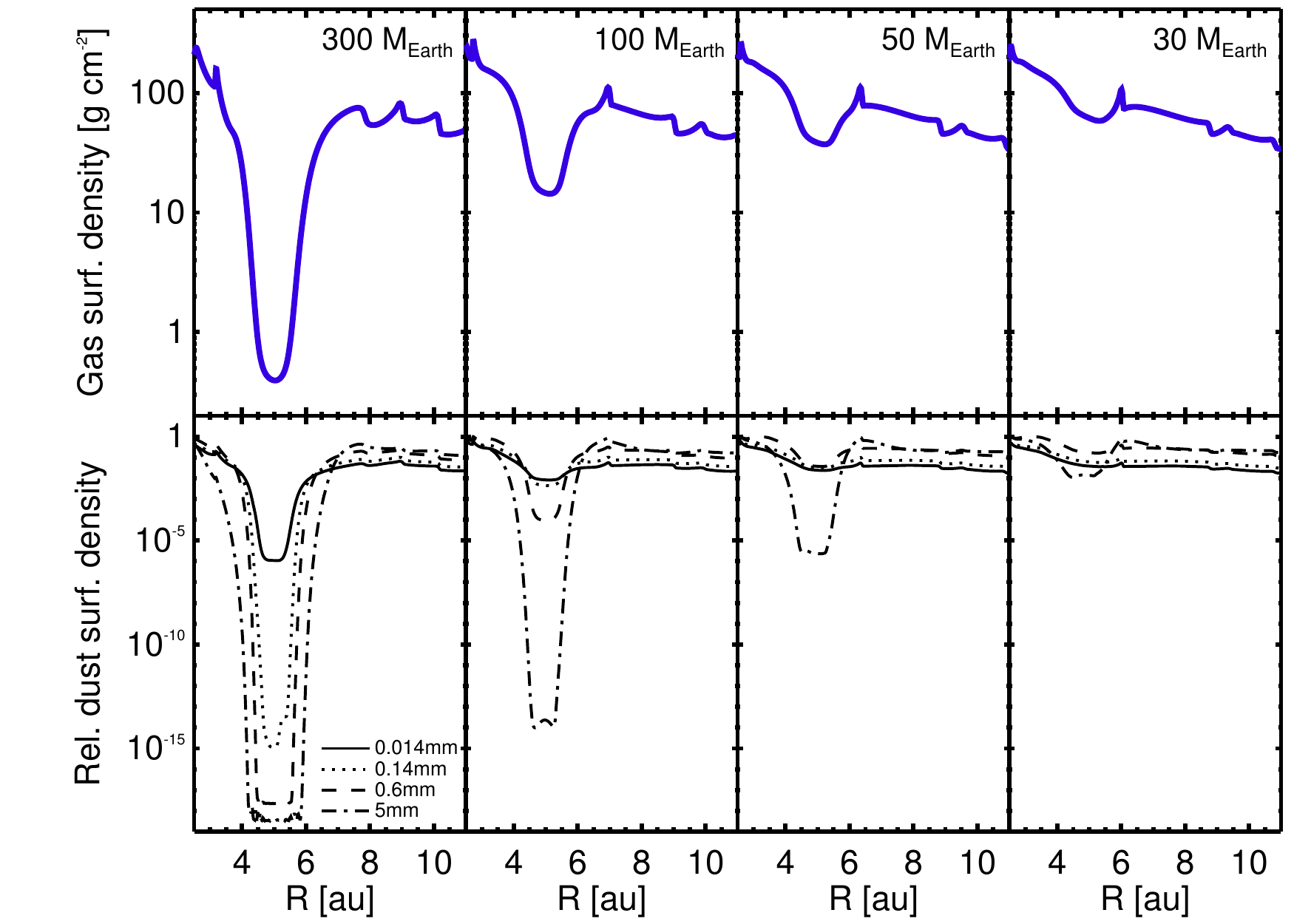} 
\caption{Top row: Radial profiles for the gas surface density for disk models with a planet at 5 au from the central star (Section~\ref{sec:high_visc}). At each stellocentric radius, the surface density has been azimuthally averaged. Planet mass is as labeled in each panel. For these models the initial disk mass is $8 \times 10^{-3}~M_{\odot}$ and the viscosity parameter $\alpha = 10^{-3}$. Bottom row: Radial profiles of the dust surface density normalized to the peak value. The four lines correspond to four different grain sizes as labeled in the bottom left panel.
}
\vspace{0.7cm}
\label{fig:surfdens}
\end{figure*}

Figure~\ref{fig:surfdens} shows the azimuthally averaged radial profiles of gas and dust surface densities for models with planets with orbital radius of 5 au, and masses of 300, 100, 50, 30 $M_{\oplus}$ from left to right, respectively. In these simulations we set the initial total mass of the disk to $M_{\rm{disk}} = 8\times10^{-3}~M_{\odot}$ and the viscosity parameter $\alpha = 10^{-3}$. 
In all these simulations, the distribution of gas is significantly influenced by the gravitational torques between the planet and the disk \citep[e.g.][]{Goldreich:1980}. In particular, gas is cleared out at the orbital location of the planet, i.e. 5 au from the central star. The amount of gas depletion in the gap strongly depends on the planet mass, with a depletion factor ranging between slightly above two orders of magnitude in the 300 $M_{\oplus}$-planet case, down to a factor of $\sim 1.5$ for a planet with a mass of 30 $M_{\oplus}$. This is consistent with the scaling relation for the depth of the gap in gas created by giant planets, i.e. $\propto (M_{p}/M_{\star})^{-2.1} \alpha^{1.4} (h/r)^{6.5}$, as obtained by \citet{Fung:2014}.

The other main difference when comparing gaps produced by planets with different masses is their radial width. Fig.~\ref{fig:surfdens} shows that the gap radial width decreases with decreasing planet mass. This is in general agreement with numerical simulations of the disk-planet interaction, which predict the radial width of the gap to be proportional to the planet Hill radius, $r_{\rm{H}} = r_p (M_{p}/3M_{\star})^{1/3}$, where $r_p$ and $M_p$ are the planet orbital radius and mass, and $M_{\star}$ the stellar mass, respectively, although with different normalization factors \citep[e.g.,][]{Bryden:1999,Wolf:2007,Rosotti:2016}. 

The effect of the planet-disk interaction is even more pronounced when one considers the spatial distribution of dust particles. This is due to the mechanism of \textit{radial drift} of solids in a gaseous environment, which gives solids a component to their velocity vector toward the direction of local increase of the gas pressure field \citep{Weidenschilling:1977,Brauer:2008}. In the Epstein regime, the intensity of the radial drift velocity depends on the Stokes parameter $St = \rho_{\rm{solid}} a / \Sigma_{\rm{gas}}$, where $a$ is the grain size, and $\Sigma_{\rm{gas}}$ the local gas surface density. Whereas particles with $St << 1$ are well coupled to the gas and therefore closely follow the gas spatial distribution, solids with $St \sim 0.1 - 1$ are only marginally coupled to gas and strongly affected by the radial drift mechanism.  

In the case of our models with a gap created by a planet at 5 au, dust particles right within or beyond the planet orbital radius are subjected to strong radial drift towards the inner and outer edge of the gap, respectively, as a consequence of the local pressure gradient at the gap edges \citep[e.g.][]{Pinilla:2012}. This produces gaps in the dust component that are always more pronounced than in gas \citep[see also][]{Dipierro:2016}. 
Among the grain sizes considered in our simulations, this effect is more important for larger grains, which, at the gas densities in the gaps, have $St$-parameter values closer to $\sim 1$ than smaller grains 
(see the comparison between different lines in the bottom panels in Fig.~\ref{fig:surfdens}). Since at a given wavelength $\lambda$ the most efficient emitting grains have sizes $a \propto \lambda/2\pi$ \citep[e.g.,][]{Draine:2006,Rodmann:2006}, this effect causes the gaps in the dust continuum emission to be \textit{wider} at longer wavelengths. 

As for the dependence on planet mass, the difference between the gas and dust gaps is more pronounced for more massive planets. This is because of i) the steeper radial profile of gas density, and therefore of gas pressure as well, within the gap, which makes the radial drift of a particle of a given size more effective, ii) the larger value, closer to unity, of the $St$-parameter of dust particles within the gap, because of the lower local gas surface densities.

Besides the gap structure, another effect of the disk-planet interaction is spiral arms which are launched by the planet in the disk and can open mini-gap substructures relatively far from the planet \citep[see also][]{Bae:2017}. These substructures manifest in Fig.~\ref{fig:surfdens} as oscillations in the gas surface density radial profiles both within and beyond the gap, and are more pronounced in the case of more massive planets. Similar oscillations are seen also in the radial profile of the surface density of dust, but, as planet-induced spiral over-densities are unable to trap dust particles since they corotates with the planet \citep{Juhasz:2015}, the resultant very small amplitudes of oscillation found in our models with gas viscosity $\alpha = 10^{-3}$ make them undetectable with either ALMA or ngVLA under reasonable assumptions on the observational parameters.

\begin{figure*}
\centering
\hspace*{-1.8cm}
\includegraphics[scale=0.77,angle=270]{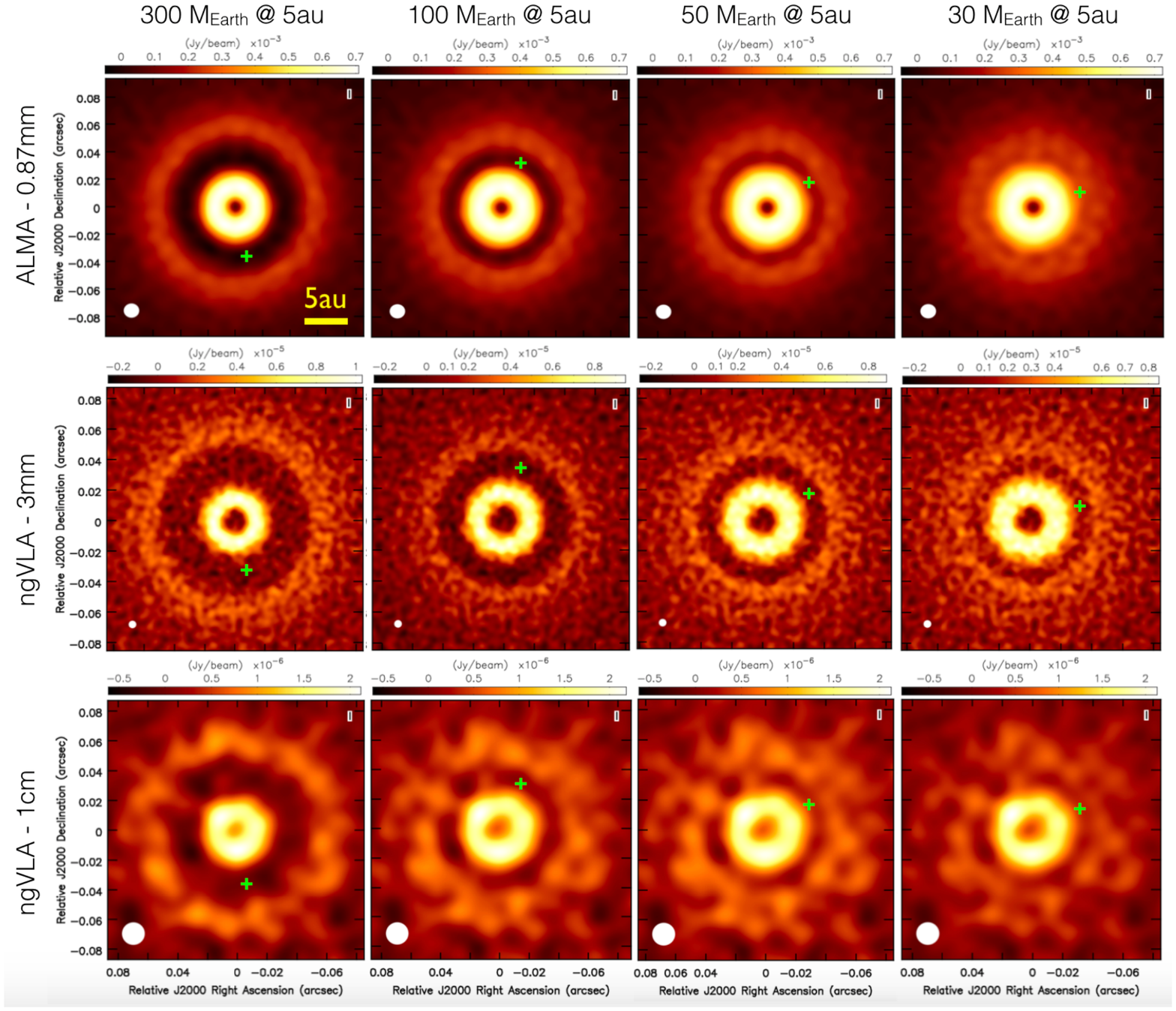} 
\caption{Simulated ALMA and ngVLA continuum maps of the disk$+$planet models with planets at 5 au from the central star (Section~\ref{sec:high_visc}). From left to right, the models contain planets with masses of $300, 100, 50, 30~M_{\oplus}$, respectively. For these models the initial disk mass is $8 \times 10^{-3}~M_{\odot}$ and the viscosity parameter $\alpha = 10^{-3}$. In each panel, the green cross indicates the location of the planet. The assumed distance is 140 pc. Top row: ALMA simulations at a wavelength of 0.87 mm. The synthesized beam, with a size of 12 mas, is shown on the bottom left corner of each panel. The $1\sigma$ rms noise on the map is 9 $\mu$Jy/beam. Center row: ngVLA simulations at a wavelength of 3 mm. The synthesized beam, with a size of 5 mas, is shown on the bottom left corner of each panel. The $1\sigma$ rms noise on the map is 0.5 $\mu$Jy/beam. Bottom row: ngVLA simulations at a wavelength of 1 cm. The synthesized beam, with a size of 16 mas, is shown on the bottom left corner of each panel. The $1\sigma$ rms noise on the map is 0.15 $\mu$Jy/beam.}
\vspace{1cm}
\label{fig:maps}
\end{figure*}

Figure~\ref{fig:maps} shows the results of the ALMA and ngVLA observations for these models. In the case of the Jupiter-mass planet at 5 au, the gap structure is clearly detected at all the three wavelengths shown here, i.e. 0.87 mm (ALMA), 3 mm and 1 cm (ngVLA). The dust emission within the gap is detected at high signal-to-noise ($> 10$), whereas the peak signal-to-noise at the gap outer edge decreases to $\approx$ 20, 7, 4 at 0.87, 3 mm, 1 cm, respectively. The decrease of the signal at larger stellocentric radii (beyond vs within the gap) is due to the decrease of both dust density and temperature. The decrease at longer wavelengths is due to the decrease of dust emissivity \citep[][]{Draine:2006}.  
Relative to ALMA at 0.87 mm, for which the beam size is only slightly smaller than the radial extent of the gap, the ngVLA at 3 mm can better spatially resolve the gap, with $\approx 5$ beams covering the gap radial width. This is due to the combined effects of better angular resolution of the ngVLA, as well as of a wider dust gap for larger grains (Fig.~\ref{fig:surfdens}, lower panels), that reflects into a wider gap for the dust emission at longer wavelengths. 

For lower mass planets the gap radial width shrinks, and ngVLA observations at 3 mm are the only ones that can spatially resolve the gap. This is true for planet masses in the range $\approx 30 - 100~M_{\oplus}$. According to our models with viscosity $\alpha = 10^{-3}$, planets with masses $\lesssim 30~M_{\oplus}$ do not produce any significant substructure that can be detected under reasonable assumptions on the observational parameters.

\begin{figure*}
\centering
\includegraphics[scale=0.65,angle=270]{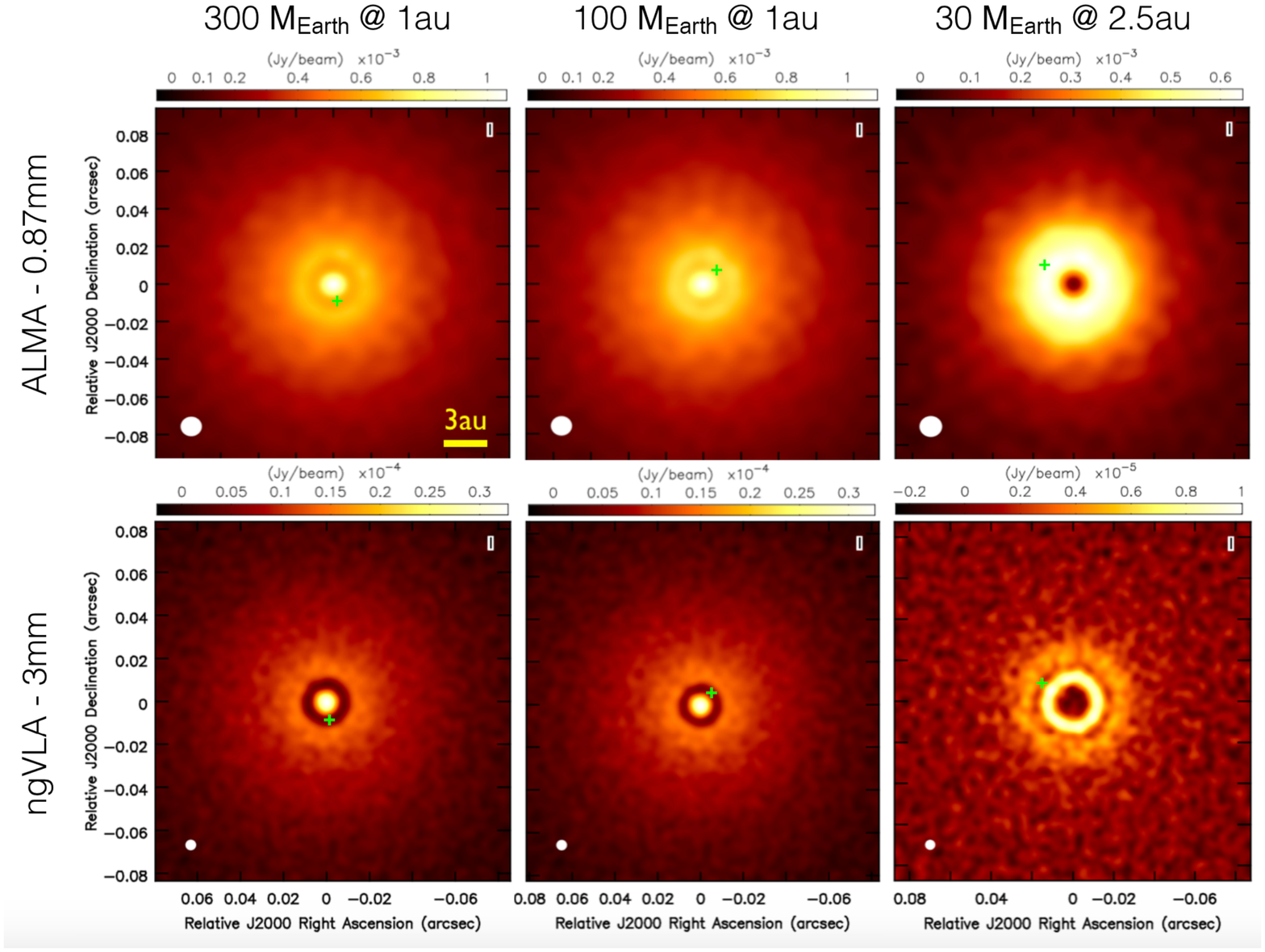} 
\caption{Simulated ALMA and ngVLA continuum maps of the disk$+$planet models with planets within 5 au from the central star (Section~\ref{sec:high_visc}). From left to right, the models contain planets with masses of $300, 100, 30~M_{\oplus}$, respectively. The planet orbital radii are 1, 1 and 2.5 au, respectively. For these models the initial disk mass is $8 \times 10^{-3}~M_{\odot}$ and the viscosity parameter $\alpha = 10^{-3}$. In each panel, the green cross indicates the location of the planet. The assumed distance is 140 pc. Top row: ALMA simulations at a wavelength of 0.87 mm. The synthesized beam, with a size of 12 mas, is shown on the bottom left corner of each panel. The $1\sigma$ rms noise on the map is 9 $\mu$Jy/beam. Bottom row: ngVLA simulations at a wavelength of 3 mm. The synthesized beam, with a size of 5 mas, is shown on the bottom left corner of each panel. The $1\sigma$ rms noise on the map is 0.5 $\mu$Jy/beam.}
\vspace{1cm}
\label{fig:maps_closein}
\end{figure*}

Figure~\ref{fig:maps_closein} shows the results for gaps created by planets at orbital radii $< 5$ au, i.e. a Jupiter-mass planet at 1 au (left column), a Saturn-mass planet at 1 au (center column), and a $30~M_{\oplus}$ planet at 2.5 au. The gaps are clearly detected with the ngVLA at 3 mm, whereas they are only barely visible in the ALMA maps at 0.87 mm for models with the Jupiter- and Saturn-mass planets,  and not detected in the case of the $30~M_{\oplus}$ planet.

\subsection{Models with $\alpha = 10^{-5}$}
\label{sec:low_visc}

\begin{figure*}
\centering
\hspace*{-1.7cm}
\includegraphics[scale=0.76,angle=270]{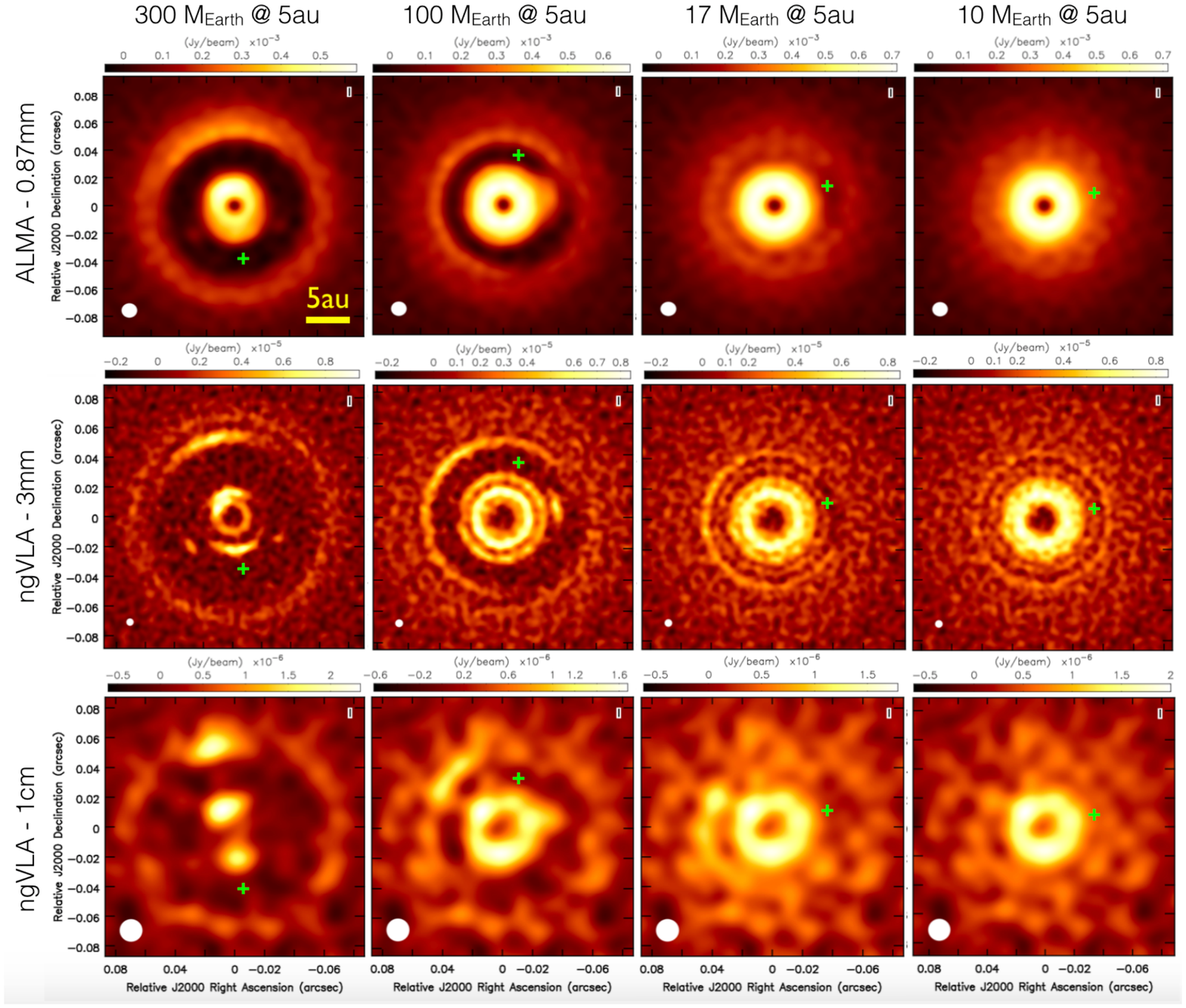} 
\caption{Same as in Figure~\ref{fig:maps} but with disk models with the viscosity parameter $\alpha = 10^{-5}$ (Section~\ref{sec:low_visc}). From left to right, the models contain planets at a distance of 5 au from the central star, and with masses of $300, 100, 17, 10~M_{\oplus}$, respectively.}
\label{fig:maps_lv}
\vspace{10mm}
\end{figure*}

Figure~\ref{fig:maps_lv} shows the results of the ALMA and ngVLA simulated observations for models with viscosity parameter $\alpha = 10^{-5}$. 
For the cases of the most massive planets considered in this study, i.e. Jupiter- and Saturn-mass planets, these maps feature strong azimuthal asymmetries in the spatial distribution of dust particles (first and second columns in Fig.~\ref{fig:maps_lv}). In particular, crescent-shaped substructures are identified both at the inner and outer gap edges. These are due to dust drifting toward the local pressure (and density) maxima produced by Rossby wave instabilities at the gap edges~\citep[][]{Lovelace:1999,Li:2000,Li:2001,Lyra:2009}. Such dust trapping mechanism may explain the azimuthal asymmetries observed in some protoplanetary disks, such as IRS 48 \citep{van der Marel:2013}. For a more thorough discussion on the detectability of vortices in disks with modelling using the LA-COMPASS code, see \citet{Huang:2018}.

The combination of higher angular resolution and more efficient dust trapping of larger particles make these asymmetries better resolved with the ngVLA at 3 mm than with ALMA. 
Furthermore, the ngVLA at 3 mm detects and spatially resolves the concentration of dust particles inside the gap at the same orbital radius of the planet. These concentrations correspond to particles located at the Lagrangian points L$_{\rm{4}}$ and L$_{\rm{5}}$ of the star-planet system~\citep[see][]{Lyra:2009}. In Fig.~\ref{fig:maps_lv} the Jupiter-mass planet is located towards the south side of the gap and is rotating counter-clockwise, so that the Lagrangian points L$_{\rm{4}}$ and L$_{\rm{5}}$ are towards the West and East directions, respectively. In the case of the Saturn-mass planet, the only detected dust concentration inside the gap is at the Lagrangian point L$_{\rm{4}}$, as in our simulations dust particles are found to leave the Lagrangian point L$_{\rm{5}}$ on a shorter timescale than L$_{\rm{4}}$.

\begin{figure*}
\centering
\includegraphics[scale=1.0]{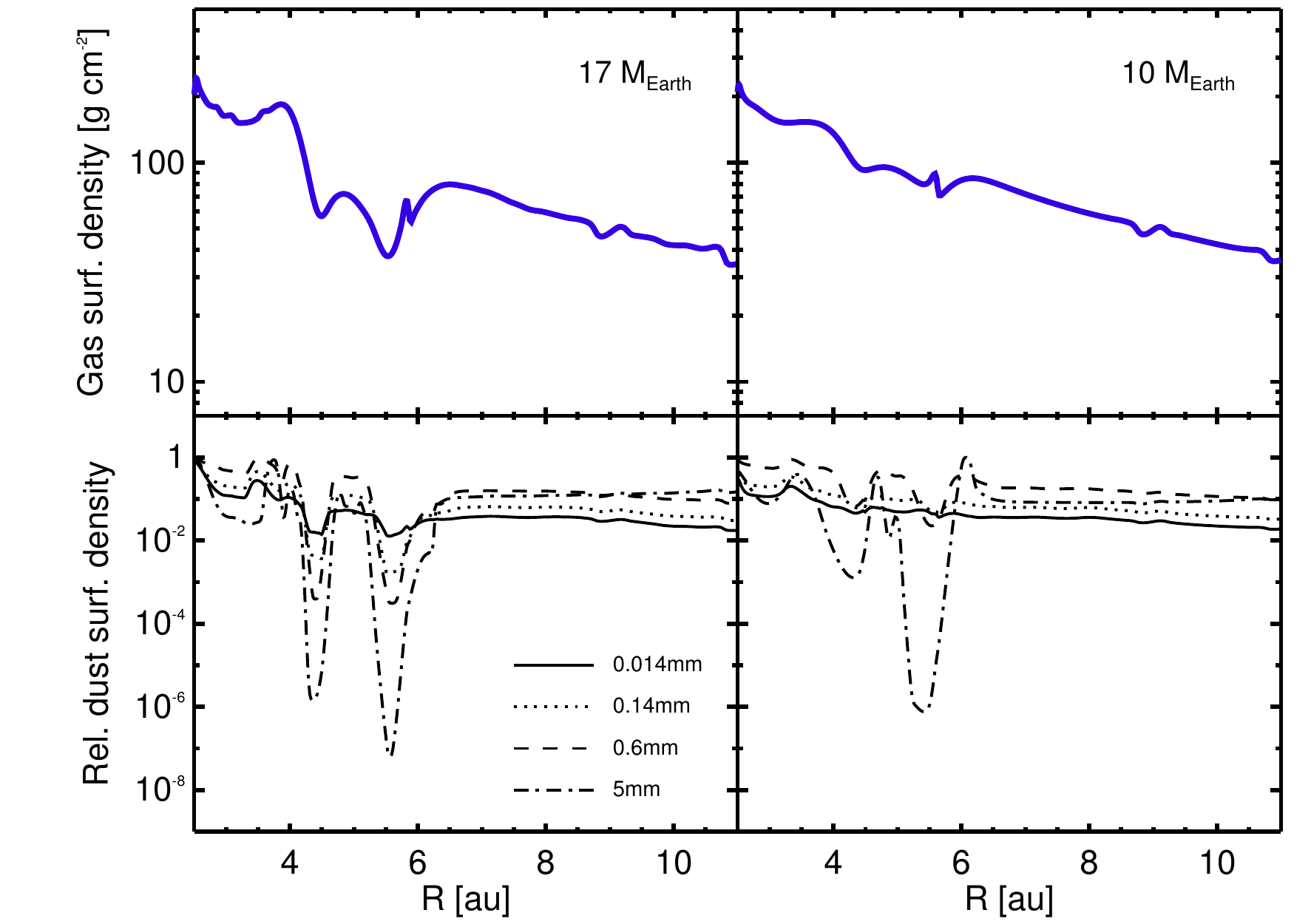} 
\caption{Same as in Figure~\ref{fig:maps} but for models with viscosity parameter $\alpha = 10^{-5}$ (Section~\ref{sec:low_visc}).}
\label{fig:surfdens_lv}
\end{figure*}

In the models with lower mass planets, i.e. 17 and 10 $M_{\oplus}$ (last two columns in Fig.~\ref{fig:maps_lv}) the ngVLA at 3 mm detects dust inside the gap and co-rotating with the planet at the orbital radius of 5 au. This feature is evident also in the azimuthally averaged radial profiles of the surface density for both gas and dust (Fig.~\ref{fig:surfdens_lv}). 
This phenomenon of a super-Earth planet opening two annular gaps to either side of its orbit was already recognized by \citet{Goodman:2001} as due to the action of Lindblad torques in disks with low viscosity \citep[$\alpha \lesssim 10^{-4}$,][]{Dong:2017}. Dust particles drift following the associated gas pressure gradients, thus producing multiple rings in the dust spatial distribution \citep{Bae:2017}.
  

\subsection{Models with lower pressure scale height}
\label{sec:low_h}

As described in Section~\ref{sec:method}, the models presented in Sections~\ref{sec:high_visc} and \ref{sec:low_visc} consider a pressure scale height given by an aspect ratio $h/r = 0.05 \times (r/5\rm{au})^{0.25}$. 
The value of the aspect ratio at the planet location is particularly important as it sets the intensity of the pressure forces which are responsible for closing the planetary gap. Higher values of the disk aspect ratio correspond to stronger pressure forces, which make more difficult for a planet of a given mass to open a gap. Conversely, disks with a \textit{lower} pressure scale height will be more prone to the formation of gaps by \textit{lower} mass planets. 

Different criteria of gap opening have been proposed in the literature and they generally predict a rather steep dependence between the mass of the lightest planet capable of opening a gap and the local aspect ratio of the disk. For example, the \textit{thermal criterion}, derived by imposing that the planet Hill radius is greater than the local pressure scale height, yields $M_p \simgreat 3 M_{\star} (h/r)^3$~\citep{Lin:1993}. A similar exponent is derived by the \textit{viscous criterion} for gap-opening, obtained by equating the timescales for gap opening and gap closing: $M_p \simgreat (27\pi/8)^{1/2} M_{\star} \alpha^{1/2} (h/r)^{5/2}$ \citep[see also][]{Dipierro:2017}.

\begin{figure*}
\centering
\hspace*{-0.5cm}
\includegraphics[angle=270,scale=0.67]{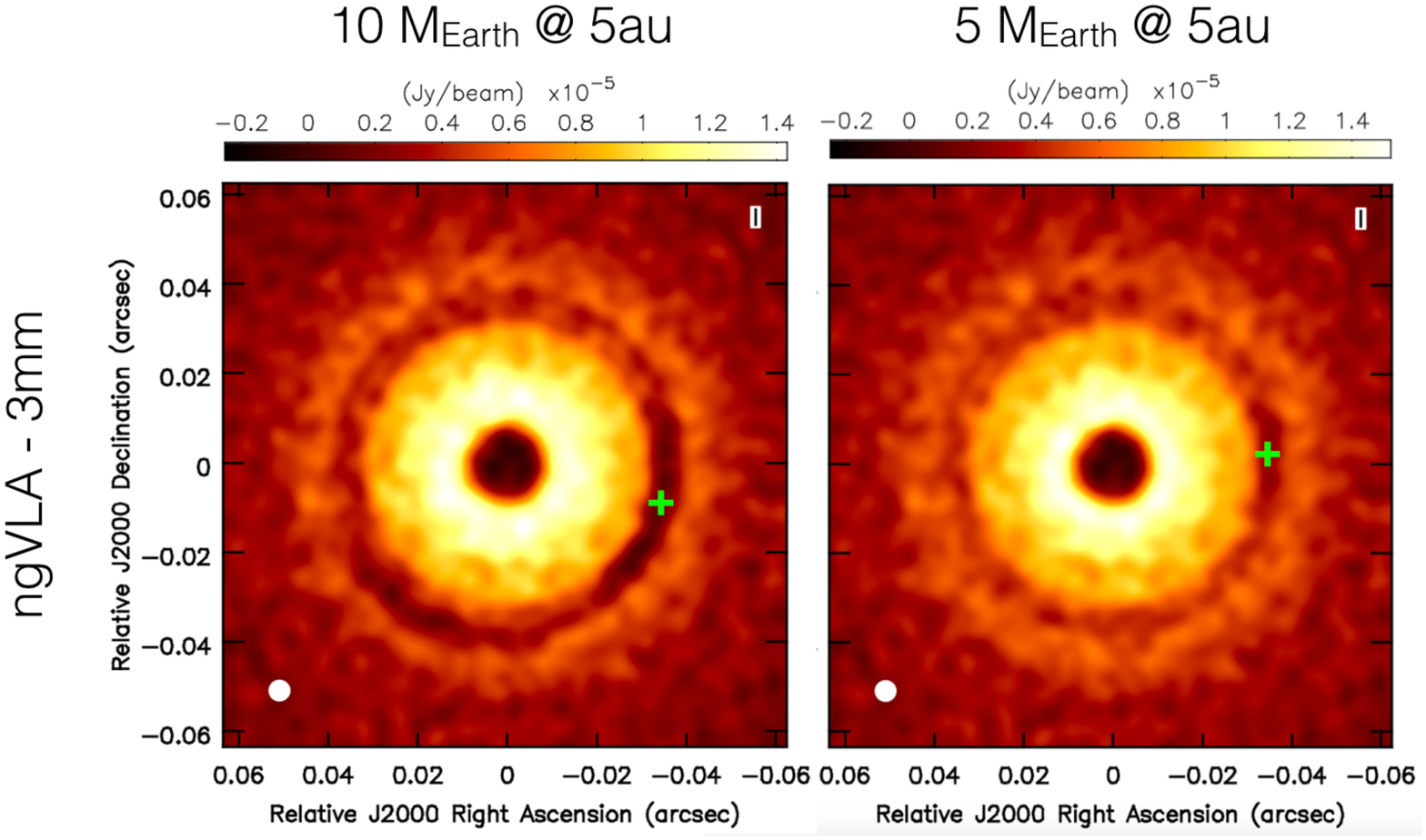} 
\caption{Simulated ALMA continuum maps at 3 mm for the case of disks with a lower pressure scale height, i.e. $h = 0.025\times(r/\rm{5au})^{1.25}$, than in the reference model (Section~\ref{sec:low_h}). Planets are located at an orbital radius of 5 au, and have a mass of 10 and 5 $M_{\oplus}$ in the left and right panel, respectively. For these models the initial disk mass is $4 \times 10^{-3}~M_{\odot}$ and the viscosity parameter $\alpha = 10^{-5}$. In each panel, the green cross indicates the location of the planet. The assumed distance is 140 pc. The synthesized beam, with a size of 5 mas, is shown on the bottom left corner of each panel. The $1\sigma$ rms noise on the map is 0.5 $\mu$Jy/beam.}
\label{fig:low_hr}
\end{figure*}

\begin{figure*}
\centering
\includegraphics[scale=0.85]{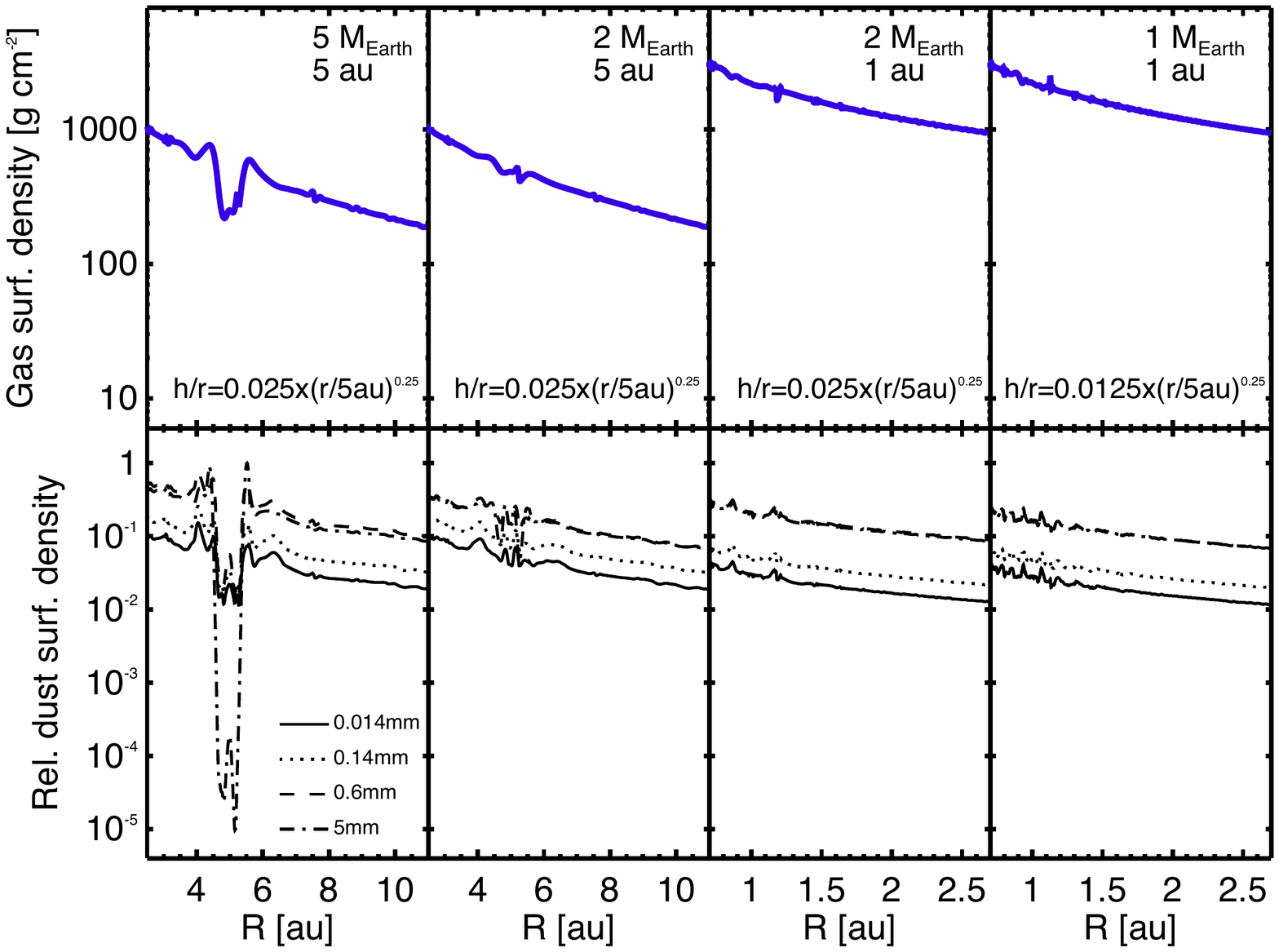}
\caption{Same as in Figure~\ref{fig:surfdens_lv} but for models with lower pressure scale heights, as labeled in the plots. The planetary mass and orbital radius are also labeled for each model (Section~\ref{sec:low_h}).}
\label{fig:surfdens_low_mass}
\end{figure*}

In order to determine the planet with the lowest possible mass that can generate an observable gap under reasonable assumptions, we ran simulations with a lower pressure scale height which is a factor of 2 lower than in our reference model, i.e. $h = 0.025\times(r/\rm{5au})^{1.25}$. In these simulations, the $\alpha$ viscosity parameter was set to 10$^{-5}$ while the initial disk mass is $M_{\rm{disk}} = 0.04~M_{{\odot}}$. The right panel in Figure~\ref{fig:low_hr} shows that, with these lower pressure forces, also a $5~M_{\oplus}$ super-Earth planet opens a gap that can be detected with the ngVLA. We also ran simulations with planets with even lower masses of 1 and $2~M_{\oplus}$ in disks with pressure scale heights a factor of 2 and 4 lower than in our reference model, but the ngVLA (and ALMA) observations resulted in no clear detection of any substructure. For this reason these maps are not shown here. The radial profiles of the gas and dust surface densities for some of these models are shown in Fig.~\ref{fig:surfdens_low_mass}.

\subsection{Different distances from Earth}
\label{sec:distance}

\begin{figure*}
\centering
\includegraphics[scale=0.8]{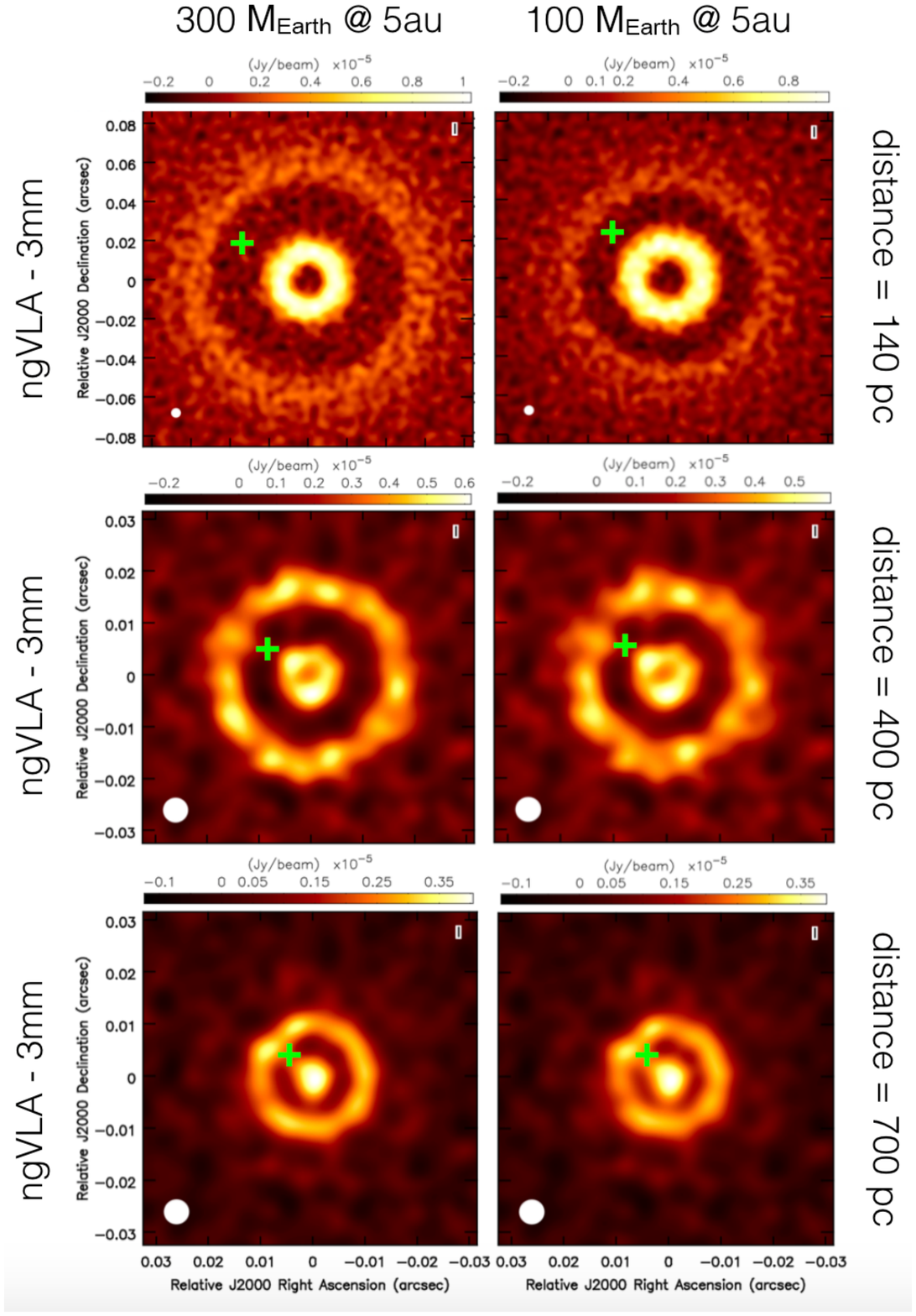} 
\caption{Simulated ngVLA continuum maps at a wavelength of 3 mm of the disk$+$planet models at different distances from the Earth. In each row, the models contain planets with masses of 300 (left column) and 100 $M_{\oplus}$ (right column), and have an orbital radius of 5 au. For these models the initial disk mass is $8 \times 10^{-3}~M_{\odot}$ and the viscosity parameter $\alpha = 10^{-3}$. In each panel, the green cross indicates the location of the planet. The synthesized beam, with a size of 5 mas, is shown on the bottom left corner of each panel. Top row: Models at a distance of 140 pc from the Earth. The $1\sigma$ rms noise on the map is 0.5 $\mu$Jy/beam. Center row: Models at a distance of 400 pc from the Earth. The $1\sigma$ rms noise on the map is 0.5 $\mu$Jy/beam. Bottom row: Models at a distance of 700 pc from the Earth. The $1\sigma$ rms noise on the map is 0.3 $\mu$Jy/beam.
}
\vspace{3mm}
\label{fig:maps_distance}
\end{figure*}

All the results shown so far for the ALMA and ngVLA simulated observations assume a distance of 140 pc, similar to the average distance of the Taurus and Ophiuchus star forming regions~\citep{Loinard:2007,Ortiz-Leon:2017}.  
Figure~\ref{fig:maps_distance} shows the results of ngVLA observations at 3 mm for models with initial disk mass $M_{\rm{disk}} = 8 \times 10^{-3}~M_{\odot}$, viscosity parameter $\alpha = 10^{-3}$, planets with an orbital period of 5 au and masses of 300 (left column) and 100~$M_{\oplus}$ (right). Top, center and bottom rows display the ngVLA continuum maps at 3 mm after adopting distances of 140, 400, and 700 pc, respectively. At these distances, the orbital radius of 5 au corresponds to angular scales of $\approx$ 36, 13 and 7 mas, respectively. 

As visible in Fig.~\ref{fig:maps_distance} the gaps created by these giant planets are detectable with the ngVLA at 3 mm up to 700 pc. At further distances these observations with an angular resolution of 5 mas do not distinguish the gap structure any longer.
A local decrease in the surface brightness can be distinguished also with ALMA at 0.87 mm for disks at a distance of 400 pc. However, due to the limitations in angular resolution, the gaps are not spatially resolved along the radial direction. For this reason, the results of the ALMA simulated observations are not shown in this figure.
 
\subsection{Rotation of the disk sub-structures}

Tracking the orbital motion of compact structures related to the planet-disk interaction, such as a circumplanetary disk \citep{Canup:2002,Zhu:2015}, or the dust particles in the Lagrangian points L$_4$ and L$_5$, would confirm their actual location in the disk, and discard the hypothesis of background or foreground sources. 

The orbital period of rotation for the planets considered in this study are of $\approx$ 11.2, 4.0, 1.0 years  for planets with orbital radii of 5, 2.5 and 1 au, respectively. Therefore, disks showing azimuthal asymmetries in the distribution of solids can reveal variations on their dust continuum emission on relatively short timescales. A planet orbiting on a circular orbit with a radius of 5 au travels at a rotational velocity of about 2.8 au yr$^{-1}$. For a face-on orientation, this corresponds to an angular velocity in the sky of 20 mas yr$^{-1}$, to be compared with the synthesized beam of 5 mas, achievable with 300 km long baselines at a wavelength of 3 mm. 

Although the disk simulations discussed in this paper are not designed to probe material immediately around the planet, and accreting onto it, ngVLA observations with the sensitivity and angular resolution presented here may detect circumplanetary disks. For example, the models by \citet{Isella:2014} predict flux densities at 3 mm of $\approx 1.9$ and 1.7 $\mu$Jy in the hot and cold start scenarios, respectively, for circumplanetary disks with a dust mass of $10^{-4}~M_{\rm{Jup}}$, outer radius of 0.5 times the planet Hill radius, and accreting at a rate of $10^{-9}~M_{\rm{Jup}}$ yr$^{-1}$ onto a Jupiter-mass planet orbiting a Solar-mass star at an orbital radius of 5 au. With the rms sensitivity considered in this work for the ngVLA simulations at 3 mm, these circumplanetary disks would be detected at $3 - 4\sigma$. For a detection at these signal-to-noise ratios, the angular precision of phased-referenced astrometry with the ngVLA would be of about 1 mas, and the proper motion due to the planet orbital rotation would be detected on a timespan of $\approx 3$ weeks. A timespan of about 1 week would be enough to observe the proper motion of a circumplanetary disk detected with the same signal-to-noise ratio but at an orbital radius of 1 au. Similar predictions for the expected circumplanetary disk fluxes at these wavelengths are derived from the $\alpha-$models of \citet{Zhu:2017}, whereas significantly brighter fluxes are obtained if adiabatic compression due to the accretion process is the main heating mechanism in these systems \citep{Szulagyi:2016,Szulagyi:2017}. Brighter circumplanetary disks, up to $\sim 100~\mu$Jy at 3 mm, are expected also around planets further from the star, where the Hill radius, and therefore the expected disk radius as well, are larger \citep{Zhu:2017}. 

The detection of orbital proper motion would be key to confirm the membership of the detected source to the stellar system, thus excluding the hypotheses of a foreground or background object. The characterization of the orbital parameters can be used to obtain an accurate dynamical estimate for the mass of the central star (or brown dwarf), while possible deviations from the local keplerian rotation velocity for some of the disk substructures may provide further evidence for the dynamical interaction with a companion in the system \citep[e.g.,][]{Regaly:2011}.

\label{sec:rotation}

\section{Discussion}
\label{sec:discussion}

The results outlined in Section~\ref{sec:results} demonstrate that the ngVLA has the potential to detect gaps and azimuthal asymmetries due to the interaction between the disk material and planets with masses down to $\sim 5 - 10~M_{\oplus}$ in the inner regions of disks located in the closest star forming regions (distance $\approx 140$ pc). 

Extending the capabilities of millimeter-wave interferometers to identify gaps opened by super-Earth/mini-Neptune planets would be highly valuable given the higher occurrence rate estimated for planets of these masses than for planets more massive than Neptune around Main Sequence stars~\citep[e.g.][]{Howard:2010,Petigura:2013a,Petigura:2013b}.

In the case of gaps opened by giant planets with $M_{p} \simgreat 100~M_{\oplus}$ at an orbital radius of 5 au, gaps can be detected in disks as far as $\sim$ 700 pc.
This would allow the investigation of disk substructures due to the disk-planet interaction in several star forming regions located within $\approx$ 700 pc from the Earth, and visible from the Northern Hemisphere. For example, the Taurus-Auriga star forming region, located at an average distance of $\approx 140$ pc~\citep{Loinard:2007,Torres:2009,Torres:2012}, contains a few hundred low-mass PMS stars. \citet{Andrews:2013} performed a survey of 227 known young stars surrounded by disks in this region with the Submillimeter Array (SMA) at a wavelength of 1.3 mm. About 35$\%$ of these sources, i.e. 80 of them, have a flux density at 1.3 mm above 10 mJy, which is very close to the flux density value of our models shown in Figures~\ref{fig:surfdens}-\ref{fig:surfdens_lv}, at a distance of 140 pc. Another region with similar numbers of disk-bearing PMS stars and at a similar distance is Ophiuchus. Dozens of disks are known with a measured flux density at 1.3 mm above 10 mJy \citep{Andrews:2007,Cox:2017}, but a complete deep survey at millimeter wavelengths has yet to be performed in this region. 

Thousands of disk-bearing PMS stars are present in the Orion molecular cloud complex at distances of $\sim 400$ pc, the richest young stellar cluster being the Orion Nebula Cluster~\citep{DaRio:2010,DaRio:2016}. So far, only a relatively small portion of the ONC, mostly toward the Trapezium Cluster, has been mapped with deep observations at millimeter wavelengths. These observations have already detected about 50 disks with fluxes close to, or greater than that of our models shown in Fig.~\ref{fig:maps_distance} at a distance of 400 pc \citep[][]{Mann:2014,Eisner:2016}. For regions at similar distances, recent ALMA and SMA observations at about 1.3 mm towards the $\sigma$ Orionis \citep{Ansdell:2017} and Serpens \citep{Law:2017} young clusters have identified 20 and 13 disks, respectively, which are brighter than our models.
Even more disks are potentially observable in the Cepheus OB young associations, located at distances of $\sim 600 - 800$ pc \citep[e.g.,][]{Kun:2008}.   

The population of proto-planets still embedded in disks surrounding PMS stars is the precursor of the population of exo-planets that is being discovered around the more mature and diskless Main Sequence stars. When a proto-planet is still embedded in its parental disk, the interaction with the disk material is known to drive an exchange of energy and angular momentum between the disk and planet itself. Moreover, material from the disk can accrete on the proto-planet. As a consequence of these processes, both the  orbital parameters and physical properties, e.g., mass, vary with time. 

The ngVLA has the capabilities to investigate the presence of proto-planets in hundreds of disks surrounding PMS stars with different properties, and in regions with different ages as well as environments. The analysis of the disk substructures with models of planet-disk interaction such as those presented in this study has the potential to constrain some of the key properties of proto-planets, such as their mass and orbital parameters.
The statistical comparison between the properties derived for the population of proto-planets still embedded in their parental disks with those obtained for exo-planets by past and future facilities can provide key observational tests to the evolutionary models of planetary systems.

For the brightest nearby disks, that can be imaged at angular resolutions of $\sim 5 - 10$ mas with ALMA at its highest frequency bands, the multi-wavelength combination of ALMA and ngVLA can highlight the different morphologies of substructures tracing different sizes of the dust particles, and test the predictions of the evolutionary models of solids in young disks \citep[see][]{Pinilla:2012}.

\section{Conclusions}
\label{sec:conclusions}

We presented the results of LA-COMPASS hydrodynamical simulations of disks containing planets interacting with the disk gas and dust. Our simulations account for both the gravitational interaction between the planet and gas and dust in the disk, as well as the mutual aerodynamical interaction between gas and dust grains with different sizes. The main goal of this work is to explore the capabilities of current (ALMA) and future (ngVLA) telescopes at sub-mm/cm wavelengths to study planet formation. To simulate observations with the ngVLA, for which a final design has yet to be established, we considered a fiducial configuration made of 300 antennas, with 30$\%$ of the antennas within a 1 km radius and maximum baselines of 300 km. We found that an outer tapering of the interferometric visibilities measured for baselines longer than 150 km, corresponding to an angular resolution of about 5 mas at 3 mm, gives the best trade-off between angular resolution and surface brightness sensitivity for the imaging of the disk models considered in this study.  

The main results are as follows.

\begin{itemize}

\item Our analysis shows that the ngVLA can detect gaps created by planets in the inner regions of nearby young disks (stellocentric radii $< 10$ au); in the case of a gas viscosity $\alpha = 10^{-3}$, the ngVLA can spatially resolve the radial width of gaps in the dust continuum emission at $\approx$ 3 mm opened by planets down to $\approx 30~M_{\oplus}$ at 5 au from the central star in the closest star forming regions (distance $\approx 140$ pc). Similar limits are obtained for the detection of gaps created by planets at stellocentric distances of 2.5 and 1 au;

\item In the case of low viscosity ($\alpha = 10^{-5}$), the ngVLA can detect azimuthal asymmetries both outside and inside the gaps opened by giant planets with masses $\simgreat~100~M_{\oplus}$. For lower mass planets, which do not produce significant asymmetric substructures in the disk, co-rotating dust can be observed down to planetary masses $\approx 10~M_{\oplus}$ for a disk with a pressure scale height $h = 0.05 \times (r/\rm{5au})^{1.25}$, and $\approx 5~M_{\oplus}$ for a disk with a factor of 2 lower pressure scale height;  

\item ALMA can detect planets down to $\approx 20-30~M_{\oplus}$ at orbital radii $< 10$ au for the disk parameters considered in this study. However, several of the azimuthal asymmetries predicted by our disk models with low viscosity and which are visible with the ngVLA are not, or only marginally, detected by ALMA;

\item With the expected astrometric precision at 3 mm, the ngVLA can observe the proper motion of azimuthal asymmetric structures associated to the disk-planet interaction, as well as possible circumplanetary disks on timescales as short as $\approx 1$ to a few weeks for planets orbiting the star at $1 - 5$ au;

\item The ngVLA has the capabilities to detect gaps opened by giant planets at distances up to $\sim 700$ pc. From the comparison between the brightness of our disk models and those measured for young disks in nearby star forming regions, we expect that the ngVLA can apply this investigation to several hundreds young disks. The statistical comparison between the properties constrained for the population of proto-planets still embedded in their parental disks with those obtained for exo-planets by past and future facilities can provide key observational tests to the models of planet formation.

\end{itemize}
 


\acknowledgments

We want to thank Chris Carilli for assistance on the simulations of the ngVLA observations and for providing us with the ngVLA antenna position files, and Giovanni Dipierro for helpful comments on the manuscript. 
This work was supported in part by the ngVLA Community Studies program, coordinated by the National Radio Astronomy Observatory, which is a facility of the National Science Foundation operated under cooperative agreement by Associated Universities, Inc.
A.I. and L. R. acknowledge support from the NSF Grant No. AST-1535809.
A.I., S.L. and H.L. acknowledge the support from Center for Space and Earth Sciences at LANL, and H.L. acknowledges the support from LANL/LDRD program.
ALMA is a partnership of ESO (representing its member states), NSF (USA) and NINS
(Japan), together with NRC
(Canada) and NSC and ASIAA
(Taiwan) and KASI (Republic of Korea), in cooperation with the Republic of Chile. The Joint ALMA Observatory is operated by ESO, AUI/NRAO, and NAOJ. The National Radio Astronomy
Observatory is a facility of the National Science Foundation operated under cooperative agreement by Associated Universities, Inc.


\end{CJK*}
\end{document}